\documentclass[sigconf]{acmart}
\AtBeginDocument{%
  \providecommand\BibTeX{{%
    \normalfont B\kern-0.5em{\scshape i\kern-0.25em b}\kern-0.8em\TeX}}}

\setcopyright{iw3c2w3}
\copyrightyear{2021}
\acmYear{2021}
\acmDOI{10.1145/3442442.3458610}

\acmConference[WWW '21 Companion]{Companion Proceedings of the Web Conference 2021}{April 19--23, 2021}{Ljubljana, Slovenia}
\acmBooktitle{Companion Proceedings of the Web Conference 2021 (WWW '21 Companion), April 19--23, 2021, Ljubljana, Slovenia}
\acmPrice{}
\acmISBN{978-1-4503-8313-4/21/04}
\begin{document}

\title[BrFAST: a Tool to Select Browser Fingerprinting Attributes for Web Authentication]{BrFAST: a Tool to Select Browser Fingerprinting Attributes for Web Authentication According to a Usability-Security Trade-off}

\author{Nampoina Andriamilanto}
\orcid{0000-0002-0224-5664}
\email{tompoariniaina.andriamilanto@irisa.fr}
\affiliation{%
  \institution{Univ Rennes, CNRS, IRISA}
  \streetaddress{263 avenue du général Leclerc}
  \city{Rennes}
  \country{France}
  \postcode{35000}
}

\author{Tristan Allard}
\orcid{0000-0002-2777-0027}
\email{tristan.allard@irisa.fr}
\affiliation{%
  \institution{Univ Rennes, CNRS, IRISA}
  \streetaddress{263 avenue du général Leclerc}
  \city{Rennes}
  \country{France}
  \postcode{35000}
}

% \renewcommand{\shortauthors}{Andriamilanto and Allard}

\begin{abstract}
  % We put ourselves as a website manager seeking to use BFP for web authentication
  In this demonstration, we put ourselves in the place of a website manager who seeks to use browser fingerprinting for web authentication.
  % First step: choose the attributes
  The first step is to choose the attributes to implement among the hundreds that are available.
  % The attribute selection platform
  To do so, we developed BrFAST, an attribute selection platform that includes FPSelect, an algorithm that rigorously selects the attributes according to a trade-off between security and usability.
  % Parameters of BrFAST
  BrFAST is configured with a set of parameters for which we provide values for BrFAST to be usable as is.
  % The two public datasets
  We notably include the resources to use two publicly available browser fingerprint datasets.
  % Extension of BrFAST
  BrFAST can be extended to use other parameters: other attribute selection methods, other measures of security and usability, or other fingerprint datasets.
  % The visualization
  BrFAST helps visualize the exploration of the possibilities during the search of the best attribute set to use, evaluate the properties of attribute sets, and compare several attribute selection methods.
  % The scenario
  During the demonstration, we compare the attribute sets selected by FPSelect with those selected by the usual methods according to the properties of the resulting browser fingerprints (e.g., their usability, their unicity).
\end{abstract}





\maketitle

% ################################ The content ################################

% ----------------- Figure: Example of Authentication Mechanism --------------
\begin{figure}
  \centering
  \includegraphics[width=\columnwidth]{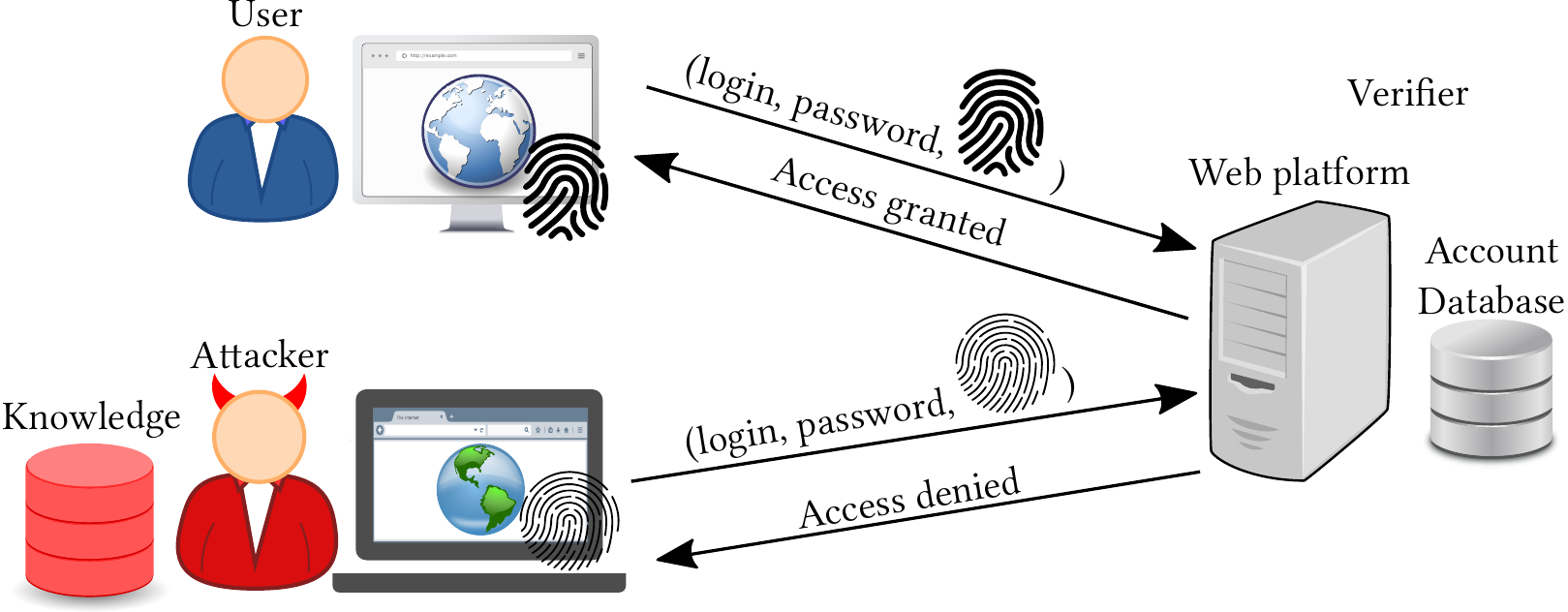}
  \caption{Example of a browser fingerprinting web authentication mechanism and a failed attack.}
  \label{fig:architecture-and-example}
  \Description[
    Example of a browser fingerprinting web authentication mechanism and a failed attack.
  ]{
    Example of a browser fingerprinting web authentication mechanism and a failed attack.
  }
\end{figure}
% --------------- End Figure: Example of Authentication Mechanism ------------

\section{Introduction}
  % What is browser fingerprinting
  Browser fingerprinting~\cite{ECK10, LRB16, GLB18, AAL20} is the collection of attributes from a web browser to build a potentially unique fingerprint.
  % Can be used for web authentication
  Initially used to track users on the web, this technique can also supplement passwords as an additional web authentication factor as depicted in Figure~\ref{fig:architecture-and-example}.
  % A lot of attributes are available
  Hundreds of attributes are available but collecting all of them is unrealistic as their usability cost (e.g., their collection time) would be too high~\cite{AALG21}.
  % Problem of correlations
  Moreover, the attributes can be correlated with each other as depicted in Table~\ref{tab:fingerprint-dataset-example}.
  % State of the art and correlation between the attributes
  Previous studies consider a small set of usual attributes~\cite{ECK10, LRB16, GLB18}, iteratively pick the attribute of the highest entropy -- or conditional entropy -- until reaching a threshold~\cite{MEN11, KZW15, FE15, BRC16, HRA18, THS18}, or evaluate every possibility~\cite{FK12}.
  % Why the usual methods are weak
  However, the entropy does not consider the correlations that occur between the attributes.
  Moreover, the entropy and the conditional entropy do not capture the usability cost induced by the use of the attributes~\cite{AAL20}.
  As for the evaluation of every possibility, we emphasize that it is impractical as the number of possibilities grows exponentially with the number of attributes.
  % The demonstration that we propose
  We propose a demonstration of FPSelect~\cite{AAL20}, a rigorous approach to select a subset of the candidate attributes such that the cost of using the fingerprints is low and a minimum security level against dictionary attacks is reached.
  % The dictionary attackers are strong
  FPSelect helps to protect against strong dictionary attackers who have the knowledge of the fingerprint distribution among the protected users.
  % The method
  To do so, it explores the space of the possible attribute sets using a greedy algorithm inspired by the Beam Search algorithm~\cite{JM09BeamSearch}.
  % The scenario of this demonstration
  This demonstration illustrates how FPSelect can be used by a website manager -- the verifier -- who seeks to use browser fingerprinting as an additional web authentication factor.
  % Comparison
  We compare the attribute sets selected by FPSelect with those selected by the usual attribute selection methods according to the properties of the resulting browser fingerprints (e.g., their usability cost, their unicity).
  % The attribute selection tool
  For this end, we developed BrFAST\footnote{
    \url{https://github.com/tandriamil/BrFAST}
  }, an attribute selection tool that performs the attribute selection given a set of parameters (e.g., fingerprint dataset, selection method).
  % The users can play with the attribute selection tool
  Users can use the set of parameters that are provided with BrFAST to perform the attribute selection, or choose their own set of parameters.
  We notably provide the resources to use two publicly available browser fingerprint datasets.
  % Extension of BrFAST
  BrFAST can be extended to use other parameters: other attribute selection methods, other measures of security and usability, or other fingerprint datasets.
  % The visualization
  BrFAST helps visualize the exploration of the possibilities during the search of the best attribute sets to use, evaluate the properties of attribute sets, and compare several attribute selection methods.

% ------------------------- Table: Example of Correlation ----------------------
\begin{table}
  \centering
  \begin{tabular}{ccccc}
    \toprule
      User  & CookieEnabled & Language & Timezone & Screen \\
    \midrule
      $u_1$ & True          & fr       & -1       & 1080   \\
      $u_2$ & True          & en       & -1       & 1920   \\
      $u_3$ & True          & it       & 1        & 1080   \\
      $u_4$ & True          & sp       & 0        & 1920   \\
      $u_5$ & True          & en       & -1       & 1080   \\
      $u_6$ & True          & fr       & -1       & 1920   \\
    \bottomrule
  \end{tabular}

  \caption{
    Example of browser fingerprints shared by users.
    The \texttt{CookieEnabled} attribute provides no distinctiveness but increases the usability cost.
    The \texttt{Timezone} and the \texttt{Language} attributes are the two most distinctive attributes, but considering them both does not improve the distinctiveness compared to considering \texttt{Language} alone due to their correlation.
  }
  \label{tab:fingerprint-dataset-example}
\end{table}
%
% For your information, here are the entropy of each attribute:
%      Browsers : 2.58 bits
%      CookieEnabled : 0.0 bits
%      Language : 1.92 bits
%      Timezone : 1.25 bits
%      Screen width : 1.0 bits
%
% ----------------------- End Table: Example of Correlation --------------------

\section{FPSelect Algorithm}
  % The main innovation that we would like to present
  In this demonstration we showcase FPSelect~\cite{AAL20}, a framework to help verifiers select the browser fingerprinting attributes to design their probe.
  % The usability / security trade-off
  To do so, FPSelect performs a trade-off between the security that the attributes provide against a dictionary attacker and the usability cost that they induce.

  % --- Dictionary attack and sensitivity measure
  \subsection{Dictionary Attack and Sensitivity Measure}
    % Consider the dictionary attackers
    We consider the attackers that managed to obtain the knowledge of a fingerprint distribution (e.g., from a stolen browser fingerprint dataset).
    These attackers are able to submit a limited number of the most common fingerprints to impersonate as many users as possible.
    % The measure as the sensitivity
    Given an attribute set, we measure the reach of the attackers by the proportion of the protected users that they manage to impersonate, and call this proportion the \emph{sensitivity}.
    % Any sensitivity measure can be used
    Any sensitivity measure can be plugged in FPSelect as long as it is monotonously decreasing when the number of selected attributes increases~\cite{AAL20}.
    Indeed, adding an attribute should decrease the sensitivity if the attribute helps distinguish different browsers, or otherwise keep the sensitivity equal.

  % --- Usability cost
  \subsection{Usability Cost Measure}
    % Usability cost measure
    FPSelect also takes a usability cost measure as a parameter, which evaluates the usability cost of an attribute set.
    % Requirements of this usability cost
    Any usability cost measure can be plugged in FPSelect as long as it is strictly increasing with the number of selected attributes.
    Indeed, adding an attribute requires at least to implement its collection, store its information, and collect it from the browser.

  % --- Lattice model
  \subsection{Lattice Model and Exploration Algorithm}
\begin{figure}
    \centering
    \includegraphics[width=\columnwidth]{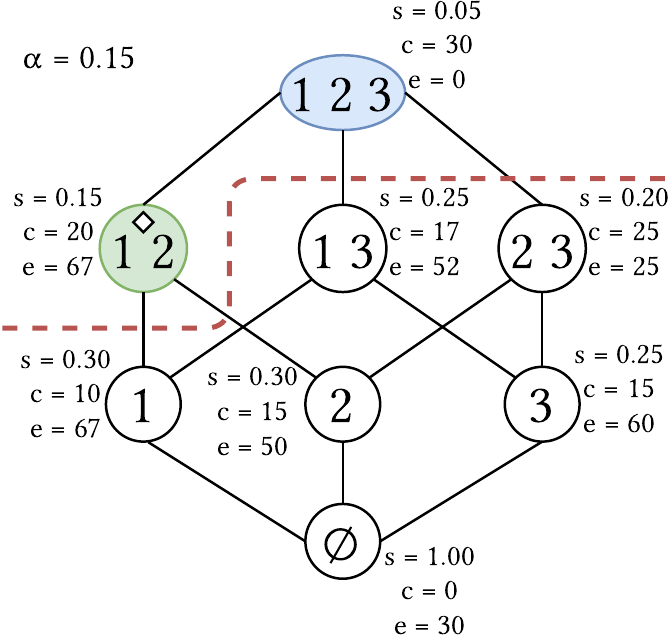}
    \caption{
      Example of a lattice of attribute sets, with their usability cost~$\mathrm{c}$, their sensitivity~$\mathrm{s}$, and their efficiency~$\mathrm{e}$.
      The sensitivity threshold is of ${\alpha = 0.15}$.
      The blue node satisfies the sensitivity threshold, the white nodes do not, and the green node with a diamond satisfies the sensitivity and minimizes the usability cost.
      The red line is the satisfiability frontier.
    }
    \label{fig:lattice-example}
    \Description[
      Example of a lattice of attribute sets, with their cost, their sensitivity, and their efficiency.
    ]{
      Example of a lattice of attribute sets, with their cost, their sensitivity, and their efficiency.
    }
    \end{figure}

    % Lattice model
    FPSelect models the possibility space as a lattice of attribute sets.
    The elements of this lattice are the subsets of the candidate attributes and the order is the subset relationship.
    % Exploration algorithm
    FPSelect leverages an exploration algorithm~\cite{AAL20} to find the attribute set that satisfies the sensitivity threshold at a low cost.
    % How the exploration works
    It starts from the empty set and explores $k$-paths in the lattice until all the paths reach the satisfiability frontier, $k$ being a parameter.
    % Satisfiability frontier
    The satisfiability frontier separates the attribute sets that satisfy the sensitivity threshold from those that do not.
    The attribute sets right above this frontier satisfy the sensitivity threshold at a lower usability cost than their supersets.
    Both the optimal solution and the solution found by FPSelect are among these attribute sets.
    % Pruning methods
    The exploration algorithm explores in priority the supersets of the most efficient\footnote{
      % How the efficiency is computed
      The efficiency of an attribute set is the ratio between the usability cost reduction and the sensitivity.
      The usability cost reduction is computed as $\text{cost}(A) - \text{cost}(C)$ with $C$ the evaluated attribute set and $A$ the candidate attributes.
    } attribute sets and includes three pruning methods~\cite{AAL20} to reduce the number of explored attribute sets.
    % Family of this algorithm
    The exploration algorithm is inspired by the Beam Search algorithm~\cite{JM09BeamSearch} and is part of the Forward Selection algorithms~\cite{SO14StepwiseRegression}.
    % Algorithm complexity
    The computational complexity of the exploration algorithm is of~${\mathcal{O}(k n^2 \omega)}$ with $n$ being the number of candidate attributes and $\omega$ being the computational complexity of the sensitivity and usability cost measures.
    The memory complexity of the exploration algorithm is of ${\mathcal{O}(k n^2)}$.
    % Reference to the lattice figure
    Figure~\ref{fig:lattice-example} shows an example of a lattice obtained from the possible attribute sets generated from three candidate attributes.

    % ------------------------ Figure: Example of a Lattice --------------------
    % ---------------------- End Figure: Example of a Lattice ------------------

  % --- Experimental results of FPSelect
  \subsection{Experimental Results}
    % Reference to our previous work showing FPSelect results
    We evaluated the performances of FPSelect and compared them to the baselines based on entropy and conditional entropy~\cite{AAL20}.
    % The experimental settings
    The experimental setting was composed of a user population of $30,000$ browsers, a number of explored paths of $1$ and $3$, a sensitivity threshold between $0.001$ and $0.025$, and a number of submissions by the dictionary attacker between $1$ and $16$.
    % The sensitivity measure
    The sensitivity was measured as the proportion of impersonated users by the most common fingerprints, considering distance functions between attributes to allow small changes.
    % The usability cost measure
    The usability cost was measured as the weighted sum between the average fingerprint size, the average fingerprint collection time, and the proportion of attribute changes among the observed consecutive fingerprints.
    % The overall results
    Compared to the attribute sets found by the baselines, those found by FPSelect generate fingerprints having a size $12$ to $1,663$ times lower, a collection time $9$ to $32,330$ times lower, and $4$ to $30$ times less attribute changes between the consecutive fingerprints.
    % The computation cost is also higher
    Although FPSelect explores three orders of magnitude more attribute sets compared to the baselines, the usability cost reduction is reflected on each authentication performed by each user.

\section{Attribute Selection Tool}
  % The demonstration platform as a basis of the demonstration: BrFAST
  We have implemented FPSelect and wrapped it into a full-fledged attribute selection tool: BrFAST.
  % The set of parameters
  BrFAST is configured with a set of parameters used to process the attribute selection, for which we provide values for anyone to directly use BrFAST as is.
  % Possibility of extension
  BrFAST is modular: other attribute selection methods or measure functions can be plugged-in easily.
  % Replay of execution traces
  As the attribute selection process can take time, BrFAST supports the replay of execution traces.
  % Software architecture
  We developed BrFAST as a web application in Python3, used Flask\footnote{
    \url{https://flask.palletsprojects.com}
  } for the web application, and used D3.js\footnote{
    \url{https://d3js.org}
  } for the visualization of the lattice exploration.

  \subsection{Parameters of the Attribute Selection Tool}
    \paragraph{An attribute selection method}
      The implemented attribute selection methods are the entropy and the conditional entropy, together with FPSelect which is configured with the number of paths explored in the lattice of the possibilities.

    \paragraph{A browser fingerprint dataset}
      % Assumption on the browser fingerprint dataset for the training
      The fingerprint dataset is collected from the browser population to protect with the fingerprints being composed of the complete set of attributes.
      % Dataset provided with the demonstration
      BrFAST includes the resources needed to use two publicly available browser fingerprint datasets.
      The first dataset\footnote{
        \url{https://github.com/Spirals-Team/FPStalker}
      } is a sample of the dataset used in the FPStalker study~\cite{VLRR18} and the second comes from an experimentation processed by Henning Tillmann\footnote{
        \url{https://www.henning-tillmann.de/2013/10/browser-fingerprinting-93-der-nutzer-hinterlassen-eindeutige-spuren}
      }.

    \paragraph{Sensitivity and usability cost measures}
      BrFAST includes a sensitivity and a usability cost measure inspired by~\cite{AAL20} that can be trained on the two provided fingerprint datasets.
      The sensitivity is measured by the proportion of the users that share the $k$ most common fingerprints, with $k$ a parameter set by the verifier.
      The usability captures the memory size and the instability of the generated fingerprints.

    \paragraph{A sensitivity threshold}
      The sensitivity threshold is configured by the verifier according to her security requirements.

  \subsection{Visualizations}
    % The high level idea of the visualization
    BrFAST helps understand the inner working of FPSelect, visualize the properties of the selected attributes, and compare the attribute selection methods.
    % The inner working of FPSelect
    The inner working of FPSelect is visualized by the real-time exploration of the lattice of the possibilities -- similar to Figure~\ref{fig:lattice-example} -- and the best solution currently found.
    % The properties of the selected attributes
    The properties of an attribute set include its usability cost, its sensitivity, a sample of the resulting fingerprints together with their entropy, their unicity, and their stability.
    % Comparison with the baselines
    Using the visualization of the properties of the selected attributes, BrFAST helps to compare several attribute selection methods.

\section{Scenario}
  % Demonstration as an execution of BrFAST
  In this demonstration, we showcase FPSelect by comparing its results with those of the baselines using BrFAST.
  % Set of parameters
  As the attribute selection process can take time, we will replay traces of executions on fingerprint datasets and sets of parameters.
  % Interactivity by the audience
  These traces will be available for the audience to replay them.
  Moreover, the audience can also plug fingerprint datasets, sensitivity and usability cost measures, and sets of parameters.

\section{Conclusion}
  % The verifier that seeks to use browser fingerprints as an authentication factor
  In this demonstration, we put ourselves in the place of a website manager that seeks to use browser fingerprinting as an additional web authentication factor.
  To do so, she has to choose the attributes to collect to compose the browser fingerprints.
  % The demonstration platform that we developed
  For this purpose, we developed BrFAST, an attribute selection tool that embarks the FPSelect algorithm to rigorously select browser fingerprinting attributes according to a trade-off between security and usability.
  % The comparison of FPSelect with the baseline methods
  Using BrFAST, we compare the attribute sets that are found by FPSelect and by the usual attribute selection methods, as well as the resulting browser fingerprints.

% ############################# End of The content #############################

% NOTE No acknowledgments
% \begin{acks}
% \end{acks}

\bibliographystyle{ACM-Reference-Format}
\bibliography{main}

% NOTE No appendix
% \appendix

\end{document}